\documentclass[12pt]{article}
\usepackage{authblk}
\usepackage[bookmarksnumbered, colorlinks, plainpages=false]{hyperref}
\usepackage{amsmath, amsthm, amscd, amsfonts, amssymb, graphicx, color, booktabs}
\usepackage{xspace} 

\def\thebaroffset{0.1em}
\newcommand{\offsetoverline}[2][\thebaroffset]{\kern #1\overline{\kern -#1 #2}}%

\def\lhcb   {\mbox{LHCb}\xspace}

\def\squark    {{\ensuremath{\rm{s}}}\xspace}

\def\cquark    {{\ensuremath{\rm{c}}}\xspace}

\def\bquark    {{\ensuremath{\rm{b}}}\xspace}

\def\uquark    {{\ensuremath{\rm{u}}}\xspace}

\def\Ppi         {\ensuremath{\pi}\xspace} 
\def\pion   {{\ensuremath{\Ppi}}\xspace}
\def\piz    {{\ensuremath{\pion^0}}\xspace}
\def\pip    {{\ensuremath{\pion^+}}\xspace}
\def\pim    {{\ensuremath{\pion^-}}\xspace}

\def\hadron    {{\ensuremath{\mathrm{h}}}\xspace}
\def\hp      {\ensuremath{\hadron^+}\xspace} 
\def\hm      {\ensuremath{\hadron^-}\xspace}

\def\PK      {\ensuremath{\mathrm{K}}\xspace} 
\def\kaon    {{\ensuremath{\PK}}\xspace}
\def\K      {{\ensuremath{\kaon}}\xspace}
\def\Kbar      {{\ensuremath{\overline{\kaon}}}\xspace}
\def\Kp      {{\ensuremath{\kaon^+}}\xspace}
\def\Km      {{\ensuremath{\kaon^-}}\xspace}
\def\Kpm     {{\ensuremath{\kaon^\pm}}\xspace}

\def\KS      {{\ensuremath{\kaon^0_{\mathrm{S}}}}\xspace}

\def\Kstarz  {{\ensuremath{\kaon^{*0}}}\xspace}
\def\Kstarzb {{\ensuremath{\Kbar^{*0}}}\xspace}

\def\PD      {\ensuremath{\mathrm{D}}\xspace}  
\def\Dbar    {{\ensuremath{\offsetoverline{\PD}}}\xspace}
\def\D       {{\ensuremath{\PD}}\xspace}
\def\Db      {{\ensuremath{\Dbar}}\xspace}
\def\Dz      {{\ensuremath{\D^0}}\xspace}
\def\Dzb     {{\ensuremath{\Dbar{}^0}}\xspace}
\def\Dp      {{\ensuremath{\D^+}}\xspace}
\def\Dm      {{\ensuremath{\D^-}}\xspace}

\def\DpDm    {\ensuremath{\Dp {\kern -0.16em \Dm}}\xspace}
\def\Dstar   {{\ensuremath{\D^*}}\xspace}

\def\Dstarz  {{\ensuremath{\D^{*0}}}\xspace}
\def\Dstarzb {{\ensuremath{\Dbar{}^{*0}}}\xspace}

\def\PB      {\ensuremath{\mathrm{B}}\xspace}
\def\Bu      {{\ensuremath{\B^+}}\xspace}
\def\Bub     {{\ensuremath{\B^-}}\xspace}
\def\Bp      {{\ensuremath{\Bu}}\xspace}
\def\Bm      {{\ensuremath{\Bub}}\xspace}
\def\Bpm     {{\ensuremath{\B^\pm}}\xspace}
\def\Bmp     {{\ensuremath{\B^\mp}}\xspace}
\def\Bz      {{\ensuremath{\B^0}}\xspace}
\def\Bzb     {{\ensuremath{\Bbar{}^0}}\xspace}
\def\B       {{\ensuremath{\PB}}\xspace}
\def\Bbar    {{\ensuremath{\offsetoverline{\PB}}}\xspace}
\def\Bb      {{\ensuremath{\Bbar}}\xspace}
\def\Bs      {{\ensuremath{\B^0_\squark}}\xspace}

\def\DzDzb    {\ensuremath{\Dz {\kern -0.07em \Dzb}}\xspace}
\def\DzDstarzb    {\ensuremath{\Dz {\kern -0.07em \Dstarzb}}\xspace}
\def\DstarzDzb    {\ensuremath{\Dstarz {\kern -0.07em \Dzb}}\xspace}
\def\DstarzDstarzb    {\ensuremath{\Dstarz {\kern -0.07em \Dstarzb}}\xspace}

\def\deltaKpi      {\ensuremath{\delta_{\K\pi}^\D}\xspace}

\def\f      {\ensuremath{\mathrm{f}}\xspace}

\def\CP                {{\ensuremath{C\!P}}\xspace}
\def\rKpi                {{\ensuremath{\mathrm{r}_{\Kpi}^\D}}\xspace}

\def\xD                {{\ensuremath{x}}\xspace}
\def\yD                {{\ensuremath{y}}\xspace}

\def\Kpi    {\ensuremath{\K\pi}\xspace}
\def\CFKpi    {\ensuremath{\Km\pip}\xspace}
\def\DCSKpi    {\ensuremath{\Kp\pim}\xspace}

\newcommand{\aunit}[1]{\ensuremath{\text{\,#1}}}

\newcommand{\gev}{\aunit{Ge\kern -0.1em V}\xspace}
\newcommand{\mev}{\aunit{Me\kern -0.1em V}\xspace}
 
\newcommand{\mevc}{\ensuremath{\aunit{Me\kern -0.1em V\!/}c}\xspace}
\newcommand{\gevc}{\ensuremath{\aunit{Ge\kern -0.1em V\!/}c}\xspace}
\newcommand{\mevcc}{\ensuremath{\aunit{Me\kern -0.1em V\!/}c^2}\xspace}
\newcommand{\gevcc}{\ensuremath{\aunit{Ge\kern -0.1em V\!/}c^2}\xspace}
\def\rB {\ensuremath{r_\B}\xspace}
\def\rBp {\ensuremath{r_\Bp}\xspace}
\def\rD {\ensuremath{r_\D}\xspace}

\def\rBz {\ensuremath{r_\Bz}\xspace}
\def\deltaB {\ensuremath{\delta_\B}\xspace}
\def\deltaBp {\ensuremath{\delta_\Bp}\xspace}
\def\deltaD {\ensuremath{\delta_\D}\xspace}
\def\deltaBz {\ensuremath{\delta_\Bz}\xspace}

\def\mmsq {\ensuremath{m_-^2}\xspace}
\def\mpsq {\ensuremath{m_+^2}\xspace}

\def\ci {\ensuremath{\rm{c}_i}\xspace}

\def\si {\ensuremath{\rm{s}_i}\xspace}

\def\Ki {\ensuremath{\rm{K}_i}\xspace}
\def\Kmi {\ensuremath{\rm{K}_{-i}}\xspace}

\def\Fi {\ensuremath{\rm{F}_i}\xspace}
\def\Fmi {\ensuremath{\rm{F}_{-i}}\xspace}

\def\xm {\ensuremath{x_-}\xspace}
\def\xp {\ensuremath{x_+}\xspace}
\def\xpm {\ensuremath{x_\pm}\xspace}
\def\ym {\ensuremath{y_-}\xspace}
\def\yp {\ensuremath{y_+}\xspace}
\def\ypm {\ensuremath{y_\pm}\xspace}

\usepackage[top=1in, bottom=1in, left=1in, right=1in]{geometry} % set page dimensions
\usepackage{lineno} % include line numbers 
\usepackage[style=phys,articletitle=false,biblabel=brackets]{biblatex}

\numberwithin{equation}{section}
\definecolor{email}{rgb}{0.00,0.00,0.84}
\begin{document}
\setcounter{page}{1}

\title{
\vspace{-2.7cm}

\large \bf 12th Workshop on the CKM Unitarity Triangle\\ Santiago de Compostela, 18-22 September 2023 \\ \vspace{0.3cm}
\LARGE Measurements of the CKM angle $\gamma$ and parameters related to mixing and \CP violation in the charm at \lhcb }

\author{Innes Mackay\textsuperscript{1} on behalf of the LHCb Collaboration \\ \vspace{-0.3cm}
        \textsuperscript{1}\small University of Oxford, Oxford, United Kingdom \\ \vspace{-0.5cm} }

%\dedicatory{This paper is dedicated to Professor ABCD}
\maketitle

\vspace{-1.2cm}

\begin{abstract}

A recent combination of measurements performed by the \lhcb collaboration determined that $\gamma=(63.8^{+3.5}_{-3.7})^\circ$. The fit combined the results of $\gamma$ and charm measurements, which resulted in precision improvements for the strong-phase difference between $\Dz\to\CFKpi$ and $\Dz\to\DCSKpi$ decays, \deltaKpi, and the \Dz--\Dzb mixing parameter, \yD. In addition, new \lhcb measurements of the CKM angle $\gamma$ using $\Bz\to\D\Kstarz$ and $\Bm\to\Dstar\Km$ decays with the $\D\to\KS\hp\hm$ final state (where $\hadron=\pi,\K$) are presented.

\end{abstract} \maketitle

\section{\texorpdfstring{\lhcb}{LHCb} combination of \texorpdfstring{$\gamma$}{gamma} and charm measurements}
A key goal of flavour physics is to test the unitarity of the {CKM} matrix by overconstraining the Unitarity Triangle (UT). The angle $\gamma$ is particularly interesting because it can be measured with a negligible theoretical uncertainty using tree-level decays~\cite{ref:TheoreticalGamma}. The direct measurements of $\gamma$ are compared to indirect determinations from global fits to parameters related to the other angles and sides of the UT~\cite{ref:CKMfitter2015,UTfit:2022hsi} that are measured in decays with additional loop diagrams. Discrepancies between the direct and indirect measurements of $\gamma$ would be a sign of beyond Standard Model effects.

The angle $\gamma$ is most precisely measured using $\Bm\to\D\Km$ decays, where \D is a superposition of \Dz and \Dzb mesons. The ratio of amplitudes for the two decay paths is
\begin{equation}
    \rB e^{i(\deltaB - \gamma)} = \frac{\rm{A}(\Bm\to\Dzb\Km)}{\rm{A}(\Bm\to\Dz\Km)}, 
\end{equation}
where \rB and \deltaB are the amplitude ratio and strong-phase difference between the decays. The angle $\gamma$ can be determined by exploiting interference between \mbox{$\Bm\to\Dz(\to\f)\Km$} and \mbox{$\Bm\to\Dzb(\to\f)\Km$} decays, where \f is a common final state to both \Dz and \Dzb mesons. The same equations above apply to $\Bp\to\D\Kp$ decays under the exchange $\gamma \to -\gamma$. The procedure can be generalised to \Bz and \Bs decays, because the sensitivity to $\gamma$ originates from the interference between $\bquark\to\uquark$ and $\bquark\to\cquark$ quark transitions that is shared by all \B-meson decays into a mixture of \Dz and \Dzb decays.

\begin{figure}[tb]
    \centering
    \begin{tabular}{cc}
         \includegraphics[width=6.5cm]{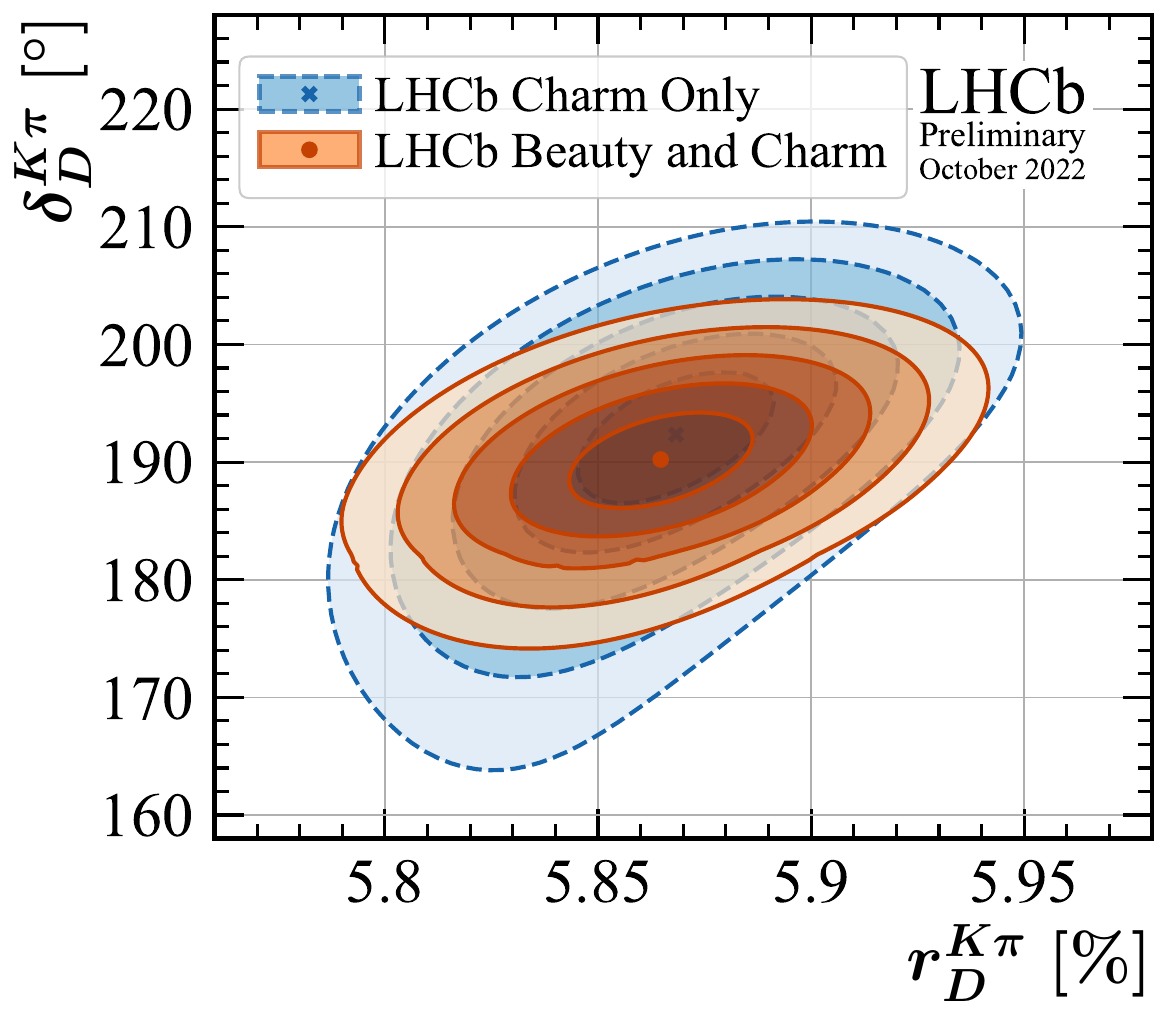} &
         \includegraphics[width=6.5cm]{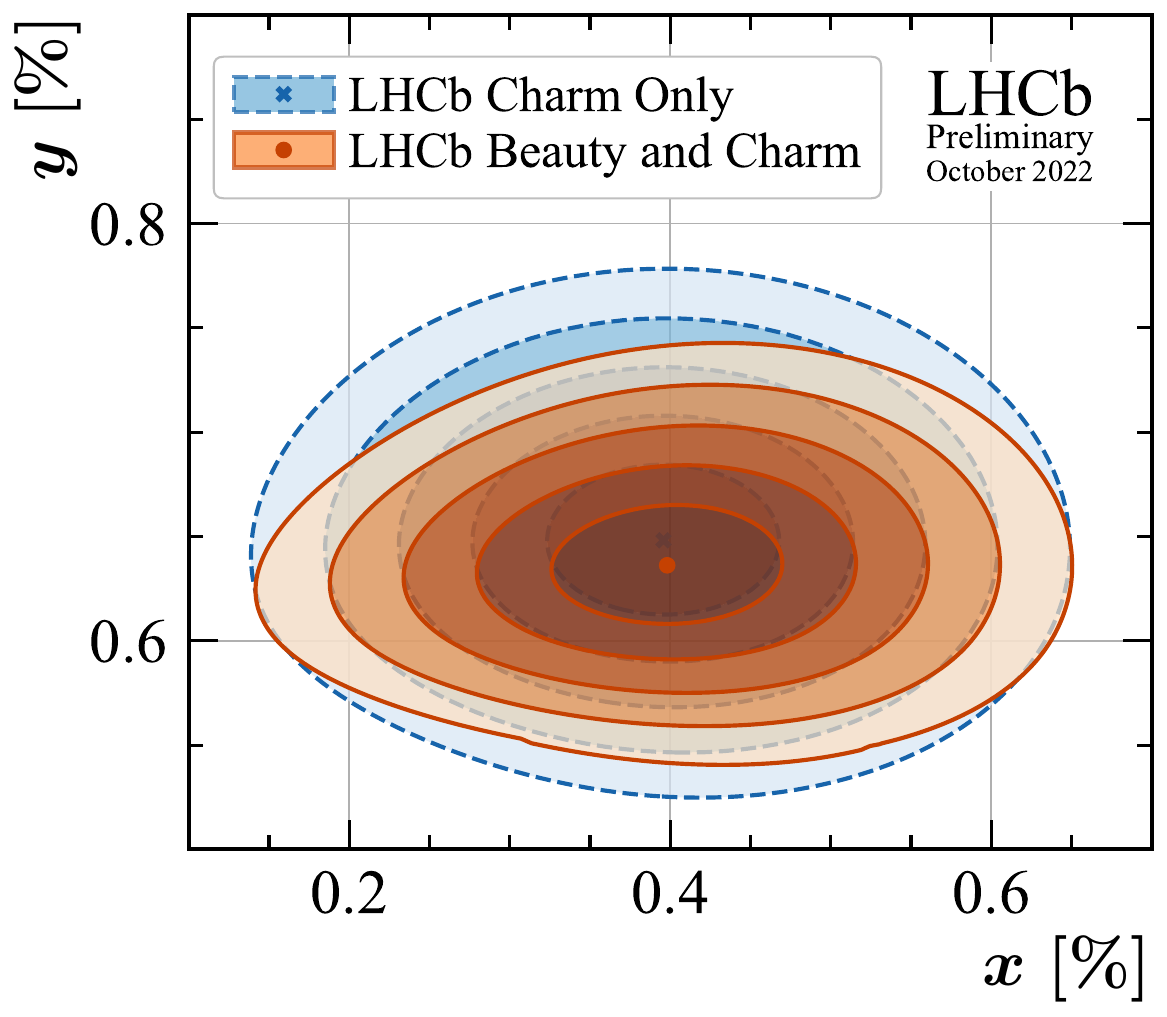} \\
    \end{tabular}
    \caption{Comparison between the 1 (68.3\%) to 5$\sigma$ confidence regions for (left) $\rKpi$ vs. $\deltaKpi$ and (right) \xD vs. \yD, as determined in a combination of \lhcb $\gamma$ and charm measurements (orange) and charm only (blue)~\cite{ref:GammaCombo}.} \label{fig:gc_deltakpi}
\end{figure}

The squared amplitude for the $\Bm\to\D(\to\f)\Km$ decay is
\begin{equation}
    |\rm{A}(\Bm)|^2 \propto \rm{A}_\D^2 + \rm{r}_\B^2\rm{A}_\Db^2 + 2\rm{A}_\D\rm{A}_\Db\rm{r}_\B\cos(\delta_\B + \delta_\D - \gamma), 
\end{equation}
where $\rm{A}_\D$ and $\rm{A}_\Db$ are the magnitude of the amplitudes for the $\Dz\to\f$ and $\Dzb\to\f$ decays, respectively, and \deltaD is the strong-phase difference between them. Therefore, external knowledge of the \D decay amplitudes is used to extract $\gamma$. The specific inputs depend on the final state of the \D decay, and are most often expressed in terms of the amplitude ratio, \rD, and strong-phase difference, \deltaD, between the $\Dz\to\f$ and $\Dzb\to\f$ decays. The reverse scenario is also possible, such that the results of the $\gamma$ analyses can be used to constrain the values of \deltaD and \rD. This is particularly useful for the $\Dz\to\CFKpi$ final state, because the determination of the \Dz--\Dzb mixing parameters, \xD and \yD, in this decay channel require knowledge of the strong-phase difference \deltaKpi~\cite{ref:LHCbMixing}. In 2021, the \lhcb collaboration combined the results of the $\gamma$ measurements with those in the charm quark sector, which determined a noticeable improvement in the precision of \deltaKpi and \yD, as is displayed in Fig.~\ref{fig:gc_deltakpi}~\cite{ref:GammaCombo}.

\begin{figure}[tb]
    \centering
    \includegraphics[width=8cm]{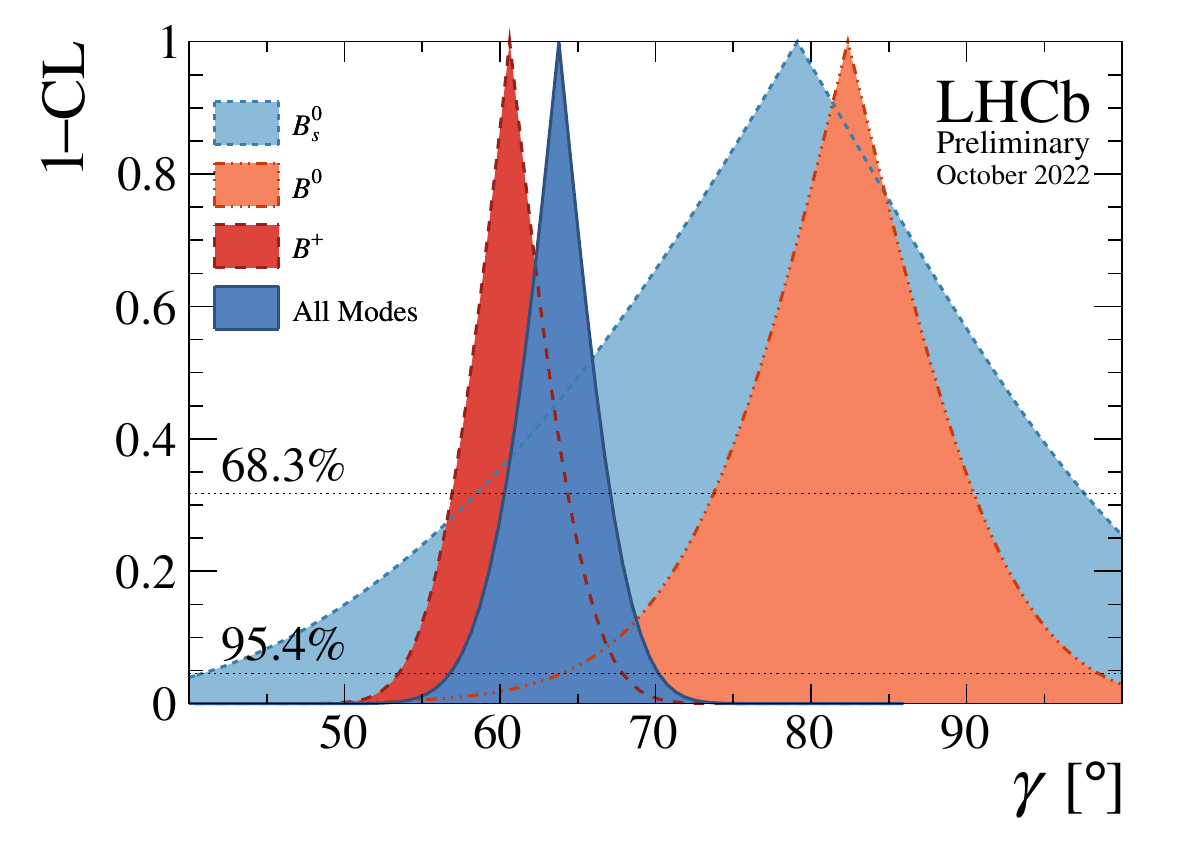}
    \caption{Confidence levels for values of the CKM angle $\gamma$ determined in a combination of $\gamma$ and charm measurements at \lhcb. The confidence levels for $\gamma$ measured using (red) \Bpm, (orange) \Bz, (light blue) \Bs and (dark blue) all \B decays are shown~\cite{ref:GammaCombo}.  } \label{fig:gc_gamma}
\end{figure}

In the combination $\gamma$ was determined to be $\gamma=(63.8^{+3.5}_{-3.7})^\circ$, which is in good agreement with the indirect measurements $\gamma=(65.6^{+0.9}_{-2.7})^\circ$~\cite{ref:CKMfitter2015} or $\gamma=(65.8\pm2.2)^\circ$~\cite{UTfit:2022hsi} depending on the statistical approach used. A comparison of $\gamma$ measurements using different \B meson decays is displayed in Fig.~\ref{fig:gc_gamma}. Those that use \Bpm mesons dominate the precision and are in minor tension with measurements made using \Bs and \Bz decays. However, recent results using time-dependent \Bs decays~\cite{ref:DsGamma} and time-independent \Bz decays~\cite{ref:myGamma, ref:AlexResult}, which both use the full \lhcb dataset, will see better agreement in the next combination. Furthermore, additional sensitivity to $\gamma$ will be achieved by including the results of two measurements made using the $\Bpm\to\Dstar\Kpm$ channel for the first time~\cite{ref:partRecoDstK,ref:fullRecoDstK}. The remainder of these proceedings will discuss some of the new \lhcb measurements of $\gamma$ that use the $\D\to\KS\hp\hm$ decay final state, where $\hadron=\pi,\K$.

\section{Measuring \texorpdfstring{$\gamma$}{gamma} with \texorpdfstring{$\D\to\KS\hp\hm$}{DtoKShh} final states}
The amplitude for the $\D\to\KS\hp\hm$ decay is phase-space dependent due to intermediate resonances. It is found that the sensitivity to $\gamma$ is optimised by studying regions of this phase space, which is categorised by the squared invariant masses of the $\KS\hp$ and $\KS\hm$ pairs, labelled \mpsq and \mmsq, respectively~\cite{ref:ggsz1, Bondar:2005ki, Bondar:2008hh, Giri:2003ty}. The phase space is divided into $2\mathcal{N}$ regions labelled from $i=-\mathcal{N}$ to $i=\mathcal{N}$ (excluding 0), and is symmetrical about the line $\mmsq = \mpsq$, where a region with $\mmsq > \mpsq$ ($\mmsq < \mpsq$) is the $i^{th}$ ($-i^{th}$) bin. For the $\D\to\KS\pip\pim$ ($\D\to\KS\Kp\Km$) decay a scheme with $\mathcal{N}=8$ ($\mathcal{N}=2$) regions is used~\cite{CLEO:2010iul}.

The signal yields in a phase-space region, $i$, are given by
\begin{align}
    \rm{N}_i(\Bm\to\D\Km) = h^{\Bm}\left[ \Fi + (\xm^2+\ym^2)\Fmi + 2\sqrt{\Fi\Fmi}\left(\ci\xm + \si\ym\right) \right],  \label{eq:bm_yield} \\
    \rm{N}_i(\Bp\to\D\Kp) = h^{\Bp}\left[\Fmi + (\xp^2+\yp^2)\Fi + 2\sqrt{\Fi\Fmi}\left(\ci\xp - \si\yp\right)\right], \label{eq:bp_yield}
\end{align}
where \xpm and \ypm are the \CP violation observables defined as $\xpm=\rB\cos(\deltaB\pm\gamma)$ and $\ypm=\rB\sin(\deltaB\pm\gamma)$. Furthermore, the $h^{\Bpm}$ are normalisation factors which absorb the detector effects and production asymmetries, the \Fi parameters are the efficiency corrected probabilities for a $\Dz\to\KS\hp\hm$ decay in a phase-space region, $i$, and the parameters \ci and \si are given by integrals over the \D decay amplitude averaged cosines and sines of the strong-phase difference, $\delta_\D$, 
\begin{align}
    \Ki &= \int_i \mathrm{d}\mmsq \mathrm{d}\mpsq\, |\mathrm{A}_\D(\mmsq, \mpsq)|^2,   \\
    \ci &= \frac{1}{\sqrt{K_i K_{-i}}}\int_{i} \mathrm{d}\mmsq \mathrm{d}\mpsq\, |\mathrm{A}_\D(\mmsq, \mpsq)||\mathrm{A}_\Db(\mmsq, \mpsq)|
    \cos\delta_\D(\mmsq,\mpsq),  \\
    \si &= \frac{1}{\sqrt{\Ki \Kmi}}\int_{i} \mathrm{d}\mmsq \mathrm{d}\mpsq\, |\mathrm{A}_\D(\mmsq, \mpsq)||\mathrm{A}_\Db(\mmsq, \mpsq)| \sin\delta_\D(\mmsq,\mpsq). 
\end{align}
The $\gamma$ measurements presented in these proceedings do not rely on amplitude models for the \ci and \si inputs. Instead, they are determined using quantum-correlated \DzDzb pairs produced at charm factories~\cite{ref:KSPiPi_SP,ref:KSKK_SP}. In each measurement, fits to the \B and \Bb meson invariant mass distributions in each phase-space region are performed with the signal yields parameterised by equations Eqs.~\ref{eq:bm_yield} and \ref{eq:bp_yield} to extract values for the \CP violation observables and thus $\gamma$, \rB and \deltaB.

\section{Measuring \texorpdfstring{$\gamma$}{g} in \texorpdfstring{$\Bz\to\D\Kstarz(892)$}{B->DKst} decays}
The measurement of $\gamma$ using $\Bz\to\D\Kstarz(892)$ decays~\cite{ref:myGamma} is performed without accounting for time dependence because the flavour of the \Bz meson at decay is unambiguously tagged by the charge on the kaon from the $\Kstarz\to\Kp\pim$ decay. Both paths to the $\Bz\to\D\Kstarz$ decay final state are colour suppressed, whereas only one is in $\Bm\to\D\Km$ decays. Therefore, the interference in $\Bz\to\D\Kstarz$ decays is expected to be around 3 times larger and thus each event has a higher sensitivity to $\gamma$ compared to $\Bm\to\D\Km$. 

\begin{figure}[tb]
    \centering
    \includegraphics[width=13cm]{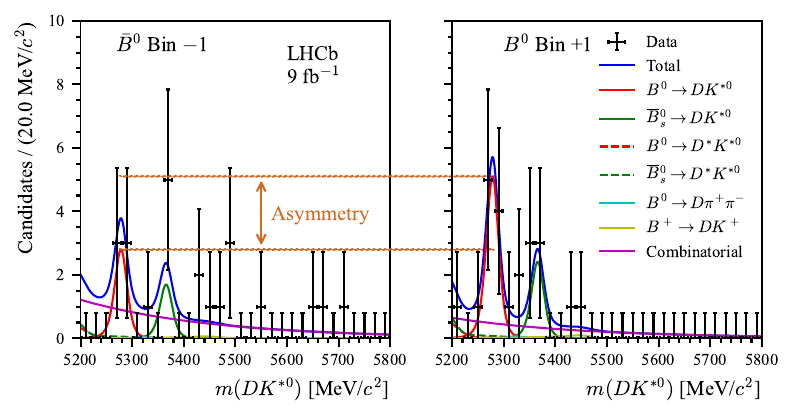} 
    \caption{Fit to the distribution of $\Bzb\to\D\Kstarzb$ candidates in the phase-space region, $i=-1$, compared to that of $\Bz\to\D\Kstarz$ decays in the phase-space region, $i=+1$. The asymmetry is highlighted on the plot.} \label{fig:b2dkst_asym}
\end{figure}

Criteria are imposed to select the $\Bz\to\D\Kstarz$ resonance in the $\Bz\to\D\Kp\pim$ phase space. However, non-signal decays still remain in this region which need to be accounted for. This is achieved by introducing a coherence factor, $\kappa=0.958^{+0.005}_{-0.046}$~\cite{ref:KappaMeasurement}, which dilutes the interference terms in Eqs.~\ref{eq:bm_yield} and ~\ref{eq:bp_yield}. In the fit to data, the \Fi values are fixed to those determined in $\Bm\to\D\pim$ decays determined in Ref.~\cite{ref:B2DK}. The relative efficiency differences across the phase space between $\Bz\to\D\Kstarz$ and $\Bm\to\D\pim$ decays are examined in simulation and found to be small. Figure~\ref{fig:b2dkst_asym} compares the fit to $\Bz\to\D\Kstarz$ candidates in the phase-space region $i=1$, with that of $\Bzb\to\D\Kstarzb$ candidates in the phase-space region $i=-1$, to provide an example of the \CP asymmetry. The analysis determines that
\begin{align}
    \gamma = (49^{+22}_{-19})^\circ, \nonumber \\
    \rBz = 0.271^{+0.065}_{-0.066}, \nonumber \\
    \deltaBz = (236^{+19}_{-21})^\circ, \nonumber
\end{align}
where the uncertainties are statistically dominated. The values presented supersede those determined using a smaller dataset in Ref.~\cite{ref:run1GammaBz}.
The measurement is required to break the four fold degeneracy of $\gamma$ values from measurements of $\Bz\to\D\Kstarz$ decays with \D decays to 2- and 4-body final states presented in Ref.~\cite{ref:AlexResult}. The combined precision is around $7^\circ$, with a best-fit value that will reduce the tensions with measurements performed using $\Bpm$ decays.

\section{Measuring \texorpdfstring{$\gamma$}{g} using \texorpdfstring{$\Bpm\to\Dstar\Kpm$}{Bpm->DstKpm} decays}
Two new measurements of $\gamma$ using $\Bm\to\Dstar\Km$ decays have recently been performed by the \lhcb collaboration. The study presented in Ref.~\cite{ref:fullRecoDstK} reconstructs the full decay chain, whilst that described in Ref.~\cite{ref:partRecoDstK} does not reconstruct the soft neutral particle from the $\Dstar\to\D X$ decay, where $X=\piz, \gamma$. There are advantages and disadvantages to both measurements. In the former, the signal yield is lower because the reconstruction efficiency of soft neutral particles is low at \lhcb; however, it has fewer backgrounds and they are easier to distinguish.

In each measurement, $\Bm\to\Dstar\pim$ decays are used as an additional signal channel. The interference in these decays is small, so they don't significantly contribute to the measurement of $\gamma$. Instead, they are used to determine the \Fi values which are simultaneously fitted alongside the \CP violation observables.

\begin{figure}[h]
    \centering
    \begin{tabular}{cc}
         \includegraphics[width=6.6cm]{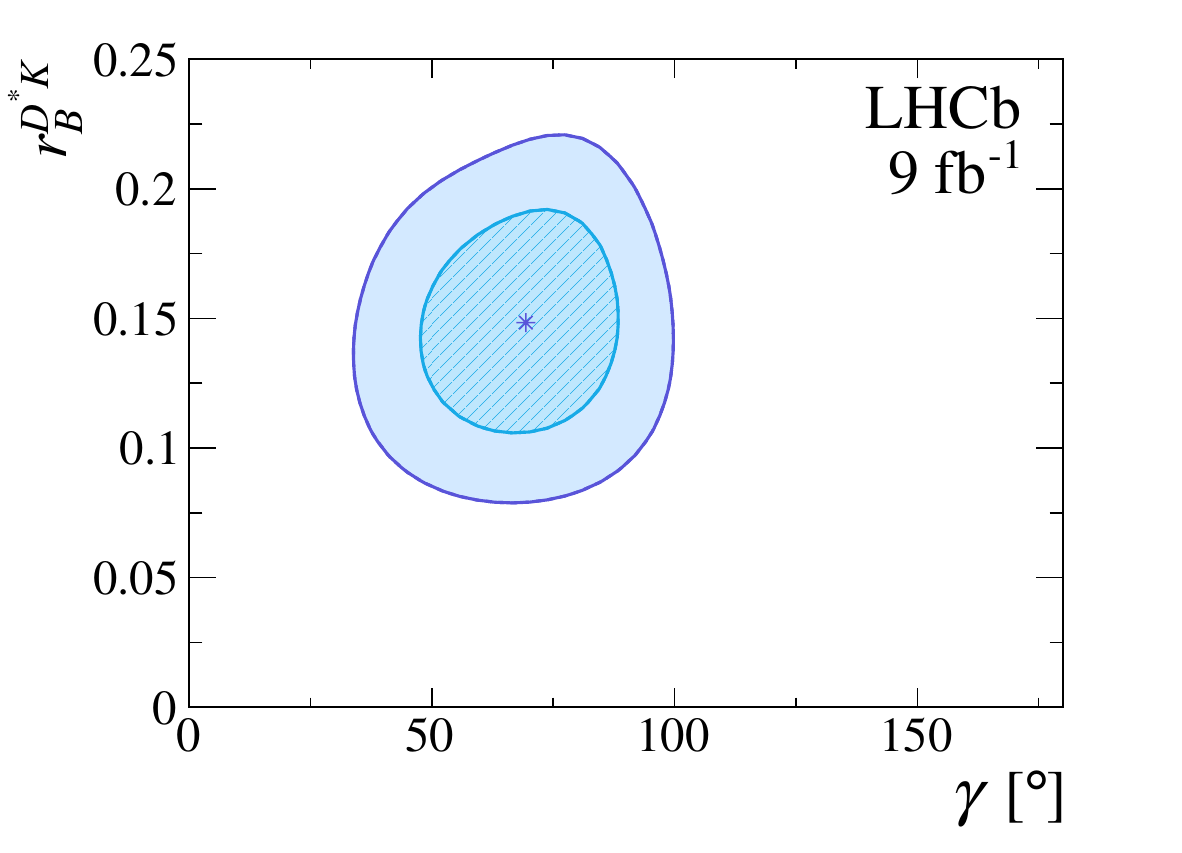} &
         \includegraphics[width=6.6cm]{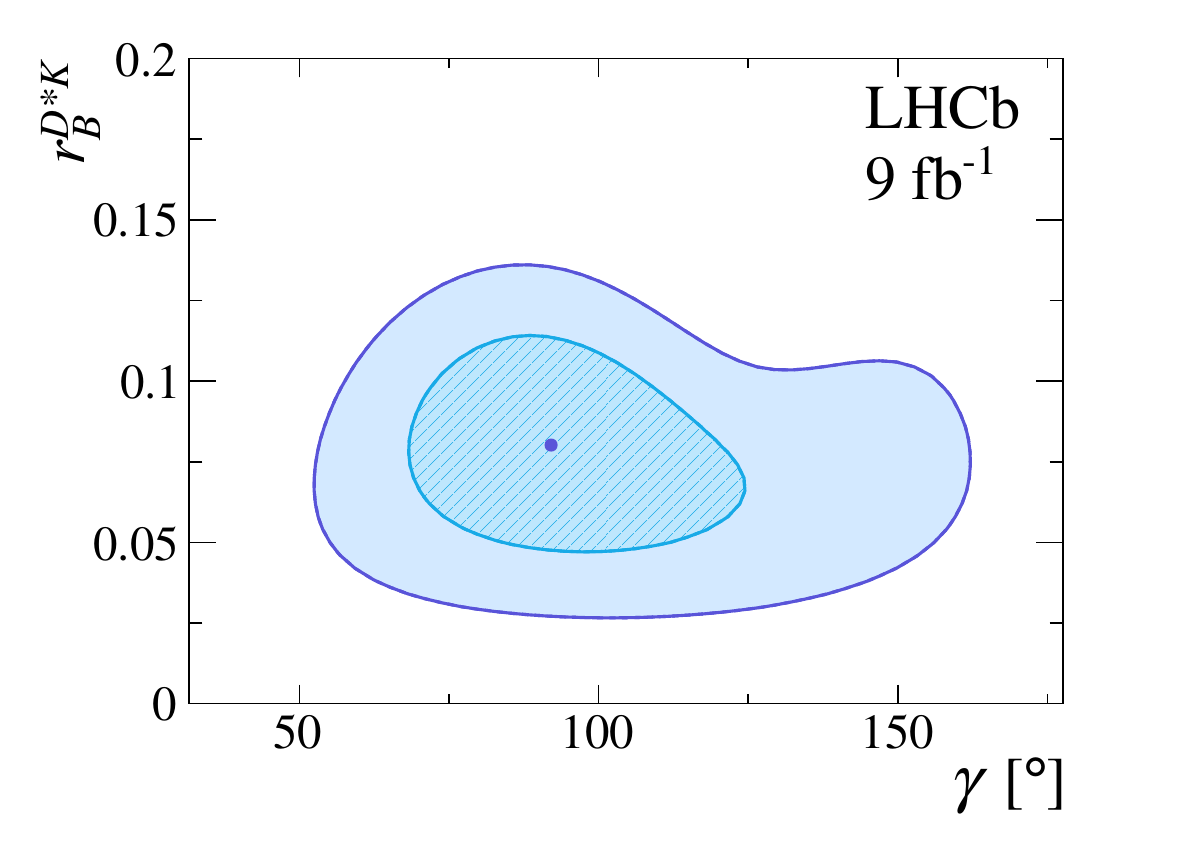} \\
    \end{tabular}
    \caption{The 68.3\% and 95.5\% confidence regions in the $\gamma$ vs. \rBz plane determined in the fits to the \CP violation observables in the study of (left) fully reconstructed and (right) partially reconstructed $\Bpm\to\Dstar\Kpm$ decays.} \label{fig:gamma_vs_rb}
\end{figure}

\begin{table}[]
    \centering
    \renewcommand{\arraystretch}{1.3}
    \begin{tabular}{|c|c|c|}
        \hline
        Parameter & Fully reconstructed & Partially reconstructed \\  
        \hline
        $\gamma$ & $(69^{+14}_{-14})^\circ$ & $(92^{+21}_{-17})^\circ$ \\
        $\rBp$ & $0.15 \pm 0.03$ & $0.080^{+0.022}_{-0.023}$ \\
        $\deltaBp$ & $(311 \pm 15)^\circ$ & $(310^{+15}_{-20})^\circ$ \\
        \hline
    \end{tabular}
    \caption{Fitted values of $\gamma$ and the hadronic parameters determined using fully and partially reconstructed $\Bmp\to\Dstar\Kpm$ decays.}
    \label{tab:dstark_results}
\end{table}

The values of $\gamma$, $\rB$ and $\deltaB$ determined in each analysis are displayed in Tab.~\ref{tab:dstark_results}. The two measurements are found to be statistically independent. The results are in agreement with each other, and with the average value of $\gamma$ determined in the \lhcb combination~\cite{ref:GammaCombo}. Figure~\ref{fig:gamma_vs_rb} displays a comparison between the 68.3\% and 95.5\% confidence regions in the $\gamma$ vs. \rBp plane in both measurements. The uncertainty on $\gamma$ is inversely proportional to the value of \rBp determined in the fit, which leads to larger confidence regions in the $\gamma$ vs. \rBp plane for the partially reconstructed analysis despite the \CP violation observables being more precise. In combinations of measurements the partially reconstructed analysis will have a larger statistical weight.

\section{Summary}
The \lhcb collaboration continues to drive precision on the CKM angle $\gamma$, with a current average value, determined in combination with charm measurements, of $\gamma=(63.8^{+3.5}_{-3.7})^\circ$~\cite{ref:GammaCombo}. New \lhcb measurements will be included in the next combination, and those that use the self-conjugate $\D\to\KS\hp\hm$, where $\hadron=\pi,\K$, final state are presented in these proceedings. An analysis of $\Bz\to\D\Kstarz$ decays determined that $\gamma=(49^{+22}_{-19})^\circ$~\cite{ref:myGamma} and will reduce tensions between measurements made using $\Bpm$ and $\Bz$ decays in the next combination. Finally, two new \lhcb measurements performed using $\Bpm\to\Dstar\Kpm$ decays~\cite{ref:partRecoDstK,ref:fullRecoDstK} for the first time provide additional sensitivity to $\gamma$, as is necessary for a more precise test of CKM unitarity.

%---------------------------------------------------------------------------------------%

%\section{References}
%\bibliography{refs} % Entries are in the refs.bib file
\printbibliography

\end{document}